\documentclass[twocolumn]{aastex62}
\usepackage{float}
\usepackage{afterpage}

\shorttitle{Draft}
\shortauthors{Sarkar et al.}
\newcommand{\jastp}{  {\it J. Atmos. Solar-Terr. Phys.}}
\begin{document}

\title{\textbf{Lorentz Force Evolution Reveals the Energy Buildup Processes during Recurrent Eruptive Solar Flares}}

\correspondingauthor{Ranadeep Sarkar}
\email{ranadeep@prl.res.in}
 \author[0000-0002-0786-7307]{Ranadeep Sarkar}
 \affil{Udaipur Solar Observatory, Physical Research Laboratory,
 Badi Road, Udaipur 313001, India} \\

 \author{Nandita Srivastava}
\affiliation{Udaipur Solar Observatory, Physical Research Laboratory,
Badi Road, Udaipur 313001, India}

 \author{Astrid M. Veronig}
\affiliation{Institute of Physics, University of Graz, A-8010 Graz, Austria}
\affiliation{Kanzelh\"ohe Observatory for Solar and Environmental Research, University of Graz, A-9521 Treffen, Austria}

\begin{abstract}
The energy release and build-up processes in the solar corona have significant implications in particular for the case of large recurrent flares, which pose challenging questions about the conditions that lead to the episodic energy release processes. It is not yet clear whether these events occur due to the continuous supply of free magnetic energy to the solar corona or because not all of the available free magnetic energy is released during a single major flaring event. In order to address this question, we report on the evolution of photospheric magnetic field and the associated net Lorentz force changes in ARs 11261 and 11283, each of which gave rise to recurrent eruptive M- and X-class flares. Our study reveals that after the abrupt downward changes during each flare, the net Lorentz force increases by $(2\textup{--}5)\times 10^{22}$ dyne in between the successive flares. This distinct rebuild-up of net Lorentz forces is the first observational evidence found in the evolution of any non-potential parameter of solar active regions (ARs), which suggests that new energy was supplied to the ARs in order to produce the recurrent large flares. The rebuild-up of magnetic free energy of the ARs is further confirmed by the observations of continuous shearing motion of moving magnetic features of opposite polarities near the polarity inversion line. The evolutionary pattern of the net Lorentz force changes reported in this study has significant implications, in particular, for the forecasting of recurrent large eruptive flares from the same AR and hence the chances of interaction between the associated CMEs.
\end{abstract}


\section{\textbf{Introduction}}
Solar flares and coronal mass ejections (CMEs) are the most energetic phenomena that occur in the solar atmosphere. Together they can release large amounts of radiation, accelerated high-energy particles and gigantic clouds of magnetized plasma that may have severe space-weather impacts \citep{Gosling,Siscoe,Daglis,2018SSRv}. Therefore, understanding the source region characteristics of these solar energetic events has become a top priority in space-science research.

Complex large active regions (ARs) on the Sun are the main sources of large flares and most energetic CMEs  \citep{Zirin,Sammis,Falconer,Wang_2008,Tschernitz,Toriumi}. Understanding the energy build-up processes in the source ARs has significant implications in particular for the case of recurrent flares, which may lead to recurrent CMEs and hence to their interaction, if the following CME has a larger speed than the preceding one.

\begin{deluxetable*}{ccccccc}[!t]

\tablecaption{Recurrent flares observed in AR 11261 and AR 11283 \label{table1}}
\tablehead{
\colhead{Active} & \multicolumn{6}{c}{Flares (GOES)}\\
\colhead{Region}&\colhead{Date} & \colhead{Start Time (UT)} &\colhead{Peak Time (UT)}  & \colhead{End Time (UT)} &\colhead{Class} &\colhead{Location}\\
\colhead{  } & \colhead{yyyy/mm/dd} & \colhead{hh:mm} &\colhead{hh:mm} & \colhead{hh:mm} &\colhead{ } &\colhead{ }
}

\startdata
 AR 11261 & 2011/08/03 & 13:17 & 13:45 & 14:30 & M6.0 & N17W30 \\
 AR 11261 & 2011/08/04 & 03:41 & 03:45 & 03:57 & M9.3 & N16W38 \\
 AR 11283 & 2011/09/06 & 01:35 & 01:50 & 02:05 & M5.3 & N13W07 \\
 AR 11283 & 2011/09/06 & 22:12 & 22:20 & 22:24 & X2.1 & N14W18 \\
 AR 11283 & 2011/09/07 & 22:32 & 22:38 & 22:44 & X1.8 & N14W31 \\
\enddata

\end{deluxetable*}

Recurrent large flares pose challenging questions regarding the conditions that lead to the episodic energy release processes \citep{Nitta,DeVore_2008,Archontis,Romano}. In particular, it is not yet clear whether these events occur due to the continuous supply of free magnetic energy to the solar corona or because not all of the available free magnetic energy is released during a single flaring event. Emergence of new magnetic flux \citep{Nitta} or photospheric shearing motions \citep{Romano} have been observed during recurrent flares. However, quantitatively it is difficult to study the
temporal evolution of the free magnetic energy of any AR due to the absence of any practical or direct method to measure the vector magnetic field in the coronal volume \citep{Wiegelmann}. Therefore, the spatial and temporal evolution of source region parameters which can be solely estimated from the photospheric magnetic field becomes important to probe the energy generation processes responsible for solar flares.

\citet{HFW} were the first to quantitatively estimate the back reaction forces on the solar surface resulting from the implosion of the  coronal magnetic field, which is required to release the energy during flares. They predicted that the photospheric magnetic fields should become more horizontal after the flare due to the act of the vertical Lorentz forces on the solar surface. 

\citet{Fisher} introduced a practical method to calculate the net Lorentz force acting on the solar photosphere. Since then, it became one of the important non-potential parameters to study the flare-associated changes in the source region characteristics. Earlier studies revealed that large eruptive flares are associated with an abrupt downward change of the Lorentz force \citep{Petrie2010,Petrie}. Comparing the magnitude of those changes associated with eruptive and confined flares, \citet{Sarkar2018} reported that the change in Lorentz force is larger for eruptive flares. However, studies on the evolution of the photospheric magnetic field and the associated Lorentz force changes for the case of recurrent eruptive large flares have not been performed so far.   

In this Letter, we study the evolution of the photospheric magnetic field and the associated net Lorentz force change during recurrent large flares which occurred in AR 11261 and AR 11283. Tracking the evolution of the net Lorentz force over the period of all the recurrent flares under study, we address the following key questions.

(i) Are the observed changes in net Lorentz force during the flare related to the linear momentum of the associated CME?

(ii) Are there any prominent signatures related to the Lorentz force evolution which might reveal the restructuring of the magnetic field after the first flare and its associated CME?
If so, these signatures might be indicative of rebuild-up of non-potentiality of the coronal magnetic field and hence the imminent more powerful flare/CME.

(iii) What causes the build-up of free magnetic energy between the successive flares? 
 
\section{\textbf{Data analysis}}\label{first_sec}
All the large recurrent M- and X-class flares that occurred in ARs 11261 (SOL2011-08-03T13:17 and SOL2011-08-04T03:41) and 11283 (SOL2011-09-06T01:35, SOL2011-09-06T22:12 and SOL2011-09-07T22:32) were well observed by the Atmospheric Imaging Assembly (AIA; \citealt{Lemen}) and the Helioseismic and Magnetic Imager (HMI; \citealt{Schou}) onboard the Solar Dynamics Observatory (SDO; \citealt{Pesnell}). To study the evolution of the photospheric magnetic field associated with the recurrent flares, we have used the HMI vector magnetogram series from the version of Space weather HMI Active Region Patches (SHARP; \citealt{Turmon}) having a spatial resolution of 0.5$''$ and 12 minute temporal cadence.

\begin{figure*}[!t]
\centering
\includegraphics[width=\textwidth]{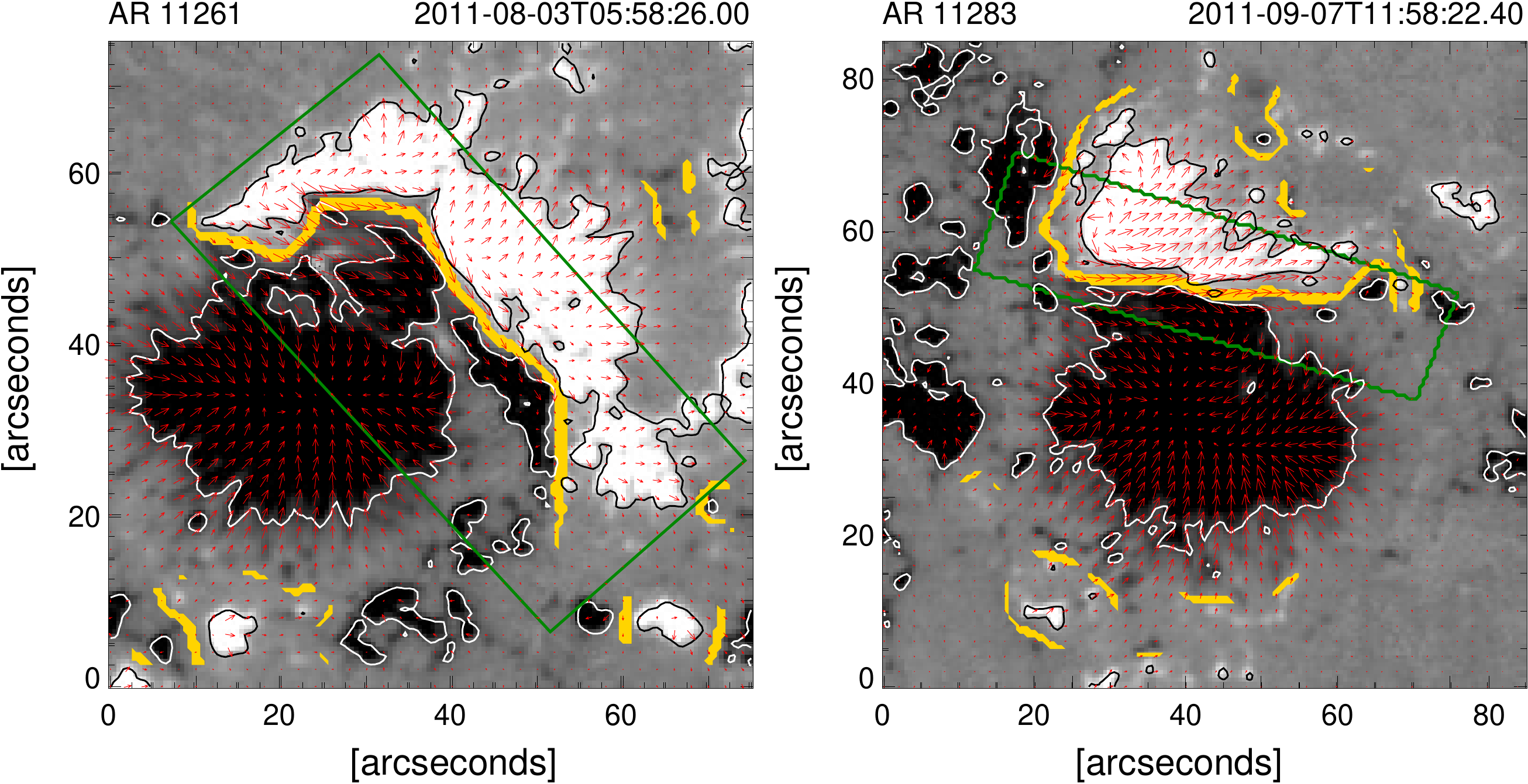}
\caption{HMI vector magnetogram of AR 11261 (left panel) and AR 11283 (right panel). The radial component ($B_r$) of the magnetic field is shown
in gray scale and the horizontal component ($B_h$) by red arrows, with saturation values $\pm 500$ G. The white/black solid line contours the region of negative/positive polarity of $B_r$ having a magnitude greater than 500 G. The green rectangular boundary encloses the selected region within which all the calculations have been done. The yellow lines illustrate the polarity inversion line.}
\label{HMI}
\end{figure*}

\begin{figure*}[!t]
\centering
\includegraphics[width=.95\textwidth]{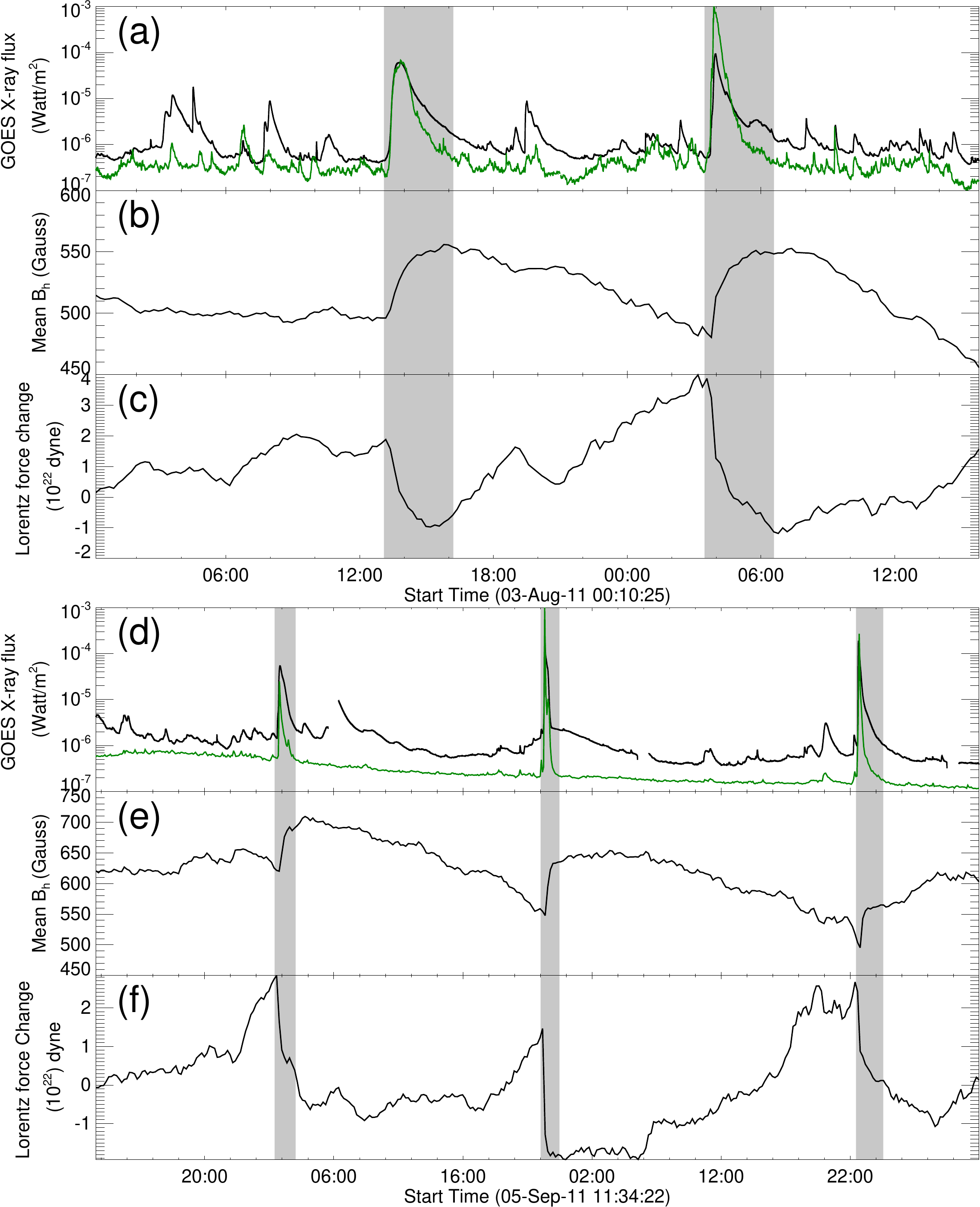}
\caption{Temporal profile of the GOES 1-8 \AA\ X-ray flux during the recurrent flares that occurred in AR 11261 (a) and AR 11283 (d). The solid green curves denote the temporal evolution of the brightening calculated within the field-of-view of the AR in the AIA 1600 \AA\ channel. Evolution of the horizontal magnetic field ((b) and (e)) and changes in the radial component of the Lorentz force ((c) and (f)) within the selected regions (shown by rectangular boxes in Figure 1) of AR 11261 and AR 11283, respectively.}
\label{force_plot}
\end{figure*}

As the errors in the vector magnetic field increase towards the limb, we have restricted our analysis to only those flares for which the flaring location of the AR was well within $\pm$ 40$^{\circ}$ from the central meridian. Moreover, we focus on the recurrent flares that initiated in the same part of the polarity inversion line of the AR and occurred within an interval of a day or less than that. This approach allows us to study the energy release and rebuild-up processes related to the recurrent flares by tracking the magnetic properties of a same flare-productive part of an AR over a period of several days. Following the aforementioned criteria, we analyze the two recurrent M-class flares (SOL2011-08-03T13:17 and SOL2011-08-04T03:41) which occurred in AR 11261 during 2011 August 3 to 4 and three recurrent flares (SOL2011-09-06T01:35, SOL2011-09-06T22:12 and SOL2011-09-07T22:32) which occurred in AR 11283 during the period 2011 September 5 to 8 (Table \ref{table1}).

To calculate the net Lorentz-force changes we have used the formulation introduced by \citet{Fisher}. The change in the horizontal and radial component of the Lorentz force within a temporal window of $\delta t$ is given as
\begin{equation}\label{first}
     \delta F_{\rm r}=\frac{1}{8\pi}\int_{A_{\rm ph}} (\delta {B_{\rm r}}^2-\delta {B_{\rm h}}^2)\\\ {\mathrm d}{\rm A} 
\end{equation}

\begin{equation}\label{second} 
   \delta F_{\rm h}=\frac{1}{4\pi}\int_{A_{\rm ph}}\delta ({B_{\rm h}}{B_{\rm r}})\\\ {\mathrm d}{\rm A}
\end{equation}   
   
   where $ B_{\rm h} $ and $ B_{\rm r} $ are the horizontal and radial components of the magnetic field, $F_{\rm h}$ and $F_{\rm r}$ are the horizontal and radial components of the Lorentz force calculated over the volume of the active region, $\rm A_{\rm ph} $ is the area of the photospheric domain containing the active region, and dA is the elementary surface area on the photosphere. Similar to \citet{Petrie}, we have reversed the signs in Equations \ref{first} and \ref{second} compared to the Equations 9 and 10 of \citet{Fisher}, as we are considering the forces acting on the photosphere from the above atmospheric volume instead of the equal and opposite forces acting on the above atmosphere from below.

As the flare related major changes in horizontal magnetic field and Lorentz forces  are expected to occur close to the polarity inversion line (PIL) \citep{Wang,Petrie2010,Petrie, Sarkar2018}, we have selected subdomains (shown by the region enclosed by the green rectangular boxes in Figure \ref{HMI}) near the PIL on the flare productive part of each AR to carry out our analysis. As the recurrent flares studied in this paper occurred from the same part of the PIL, we are able to capture the evolution of the magnetic field over several days including the time of each flares within that same selected domain on the AR. In order to define the size, orientation, and location of the selected domains we examined the post-flare loops observed in the AIA 171 and 193 \AA\ channels. Several studies have shown that the flare-reconnection process results in the simultaneous formation of a post-eruption arcade (PEA) and a flux rope above the PEA during solar eruptive events \citep{Leamon,Longcope,Qiu,Hu}. Therefore in order to capture the magnetic imprints of the recurrent large eruptive flares on the solar photosphere, we have selected our region of interest in such a way so that the major post flare arcade structures formed during each flare can be enclosed within that domain. The choice of such subdomains enables us to assume that the magnetic field on the side-boundaries enclosing the volume over those selected regions is largely invariant with time and the field strength on the top boundary is negligible as compared to that at the lower boundary on the photosphere. Therefore, only the photospheric magnetic field change contributes to the surface integrals as shown in Equations 1 and 2 to estimate the change in net Lorentz force acting on the photosphere from the above atmospheric volume.

\section{\textbf{Result and discussion}}\label{second}
\subsection{Abrupt Changes in Magnetic Field and Lorentz Force}
Figure \ref{force_plot} depicts the abrupt changes in horizontal magnetic field and the radial component of net Lorentz forces calculated within the selected region of interest as shown in Figure \ref{HMI}. The distinct changes in the magnetic properties of 
AR 11261 and AR 11283 associated with the recurrent large M- and X-class flares are discussed as follows. 
\subsubsection{Magnetic Field Evolution in AR 11261}

During the first M6.0 class flare (SOL2011-08-03T13:17) that occurred in AR 11261, the mean horizontal magnetic field increases approximately from 500 to 550 G and the associated net Lorentz force shows an abrupt downward change by approximately $2.8 \times 10^{22}$ dyne. After the M6.0 class flare the mean horizontal magnetic field started to decrease and reached about 490 G prior to the M9.3 class flare (SOL2011-08-04T03:41). During the M9.3 class flare the mean horizontal magnetic field again approximately increased to 550 G. The associated change in net Lorentz force during this flare is about $5.1 \times 10^{22}$ dyne which is almost two times larger than that associated with the previous M6.0 class flare.

In order to examine whether the kinematic properties of the associated CMEs are related to the flare induced Lorentz force changes or not, we obtain the true mass and the deprojected speed of each flare associated CME from \citet{Mishra}. The two recurrent CMEs associated with the preceding M6.0 class and the following M9.3 class flares are hereinafter referred to as CME1 and CME2, respectively. Interestingly, CME2  was launched with a speed of 1700 km s${^{-1}}$, approximately 1.5 times higher than that of CME1 (v = 1100 km s${^{-1}}$). The true masses of CME1 and CME2, estimated from the multiview of STEREO-A and -B coronagraph data, were $7.4 \times 10^{12}$ kg and $10.2 \times 10^{12}$ kg, respectively. Considering an error of $\pm$ 100 km s${^{-1}}$  in determining the CME speed \citep{Mishra} and $\pm$ 15 \% in estimating the CME mass \citep{Bein_2013,Mishra2014}, we derive the momentum of CME2 as $17 \times 10^{15} \pm 4 \times 10^{15}$ kg km s${^{-1}}$, approximately twice the momentum of CME1 ($8 \times 10^{15} \pm  2 \times 10^{15}$ kg km s${^{-1}} $). Therefore the magnitude of change in the net Lorentz force impulse during the two recurrent flares appears to be correlated with the associated CME momentum. This scenario is consistent with the flare related momentum balance condition where the Lorentz-force impulse is believed to be proportional to the associated CME momentum \citep{Fisher,ShouWang}. 

As the masses of the two CMEs were comparable, the successive Lorentz force impulse within a time window of approximately 14 hr from the same PIL of the AR with a larger change in magnitude during the following flare appears to be an important characteristic of the source AR in order to launch a high speed CME preceded by a comparatively slower one. This was an ideal condition for CME-CME interaction. Eventually, the two CMEs interacted at a distance of 145 solar radii \citep{Mishra}.

\subsubsection{Magnetic Field Evolution in AR 11283}

For all the three recurrent flares that occurred in AR 11283, the horizontal magnetic field and the net Lorentz force showed abrupt changes during each flare. It is noteworthy that the net Lorentz force increases substantially 2-4 hr prior to the occurrence of each flare, followed by a steep decrease of the same. The changes in net Lorentz force during the successive  M5.3 (SOL2011-09-06T01:35), X2.1 (SOL2011-09-06T22:12) and X1.8-class (SOL2011-09-07T22:32) flares were approximately $4 \times 10^{22}$, $3.5 \times 10^{22}$ and $3.5 \times 10^{22}$ dyne respectively. All the three flares were eruptive and the associated deprojected CME speeds were 640, 773, and 751 km s${^{-1}}$, respectively as reported in Soojeong Jang's Catalog (http://ccmc.gsfc.nasa.gov/requests/fileGeneration.php). For all the three flares the magnitude of change in net Lorentz force were almost comparable and the associated CME speeds also do not differ too much. As the three associated CMEs were launched within an interval of a day and with approximately similar speed, there was no chance of interactions among them in the interplanetary space within 1 AU. As the CDAW catalog (https://cdaw.gsfc.nasa.gov/CME\_list/) reports poor mass estimation for the aforementioned CMEs, we do not compare the linear momentum of those CMEs with the associated change in net Lorentz force.

In strong events, flare induced artifacts in the magnetic field vectors may result in magnetic transients during the stepwise changes
of the photospheric magnetic field \citep{Sun_2017}. However, these magnetic transients as reported by \citet{Sun_2017} are spatially localized in nature and temporally can be resolved within a timescale of $\approx$ 10 minutes. Moreover, the transient features do not show any permanent changes in the magnetic field evolution during the flares. The evolution of the horizontal magnetic field and the net Lorentz force as shown in Figure \ref{force_plot} are estimated within a large area on the photosphere using the 12 minute cadence vector magnetogram data. Therefore, within the time window of the stepwise changes in the horizontal magnetic field, there is no discontinuity found in the field evolution during the flares under this study as potentially occurring magnetic transients would be spatially and temporally averaged out. Hence, there are no flare related artifacts involved in the derivation of the net Lorentz force in this study.

\begin{figure*}[!t]
\centering
\includegraphics[width=\textwidth]{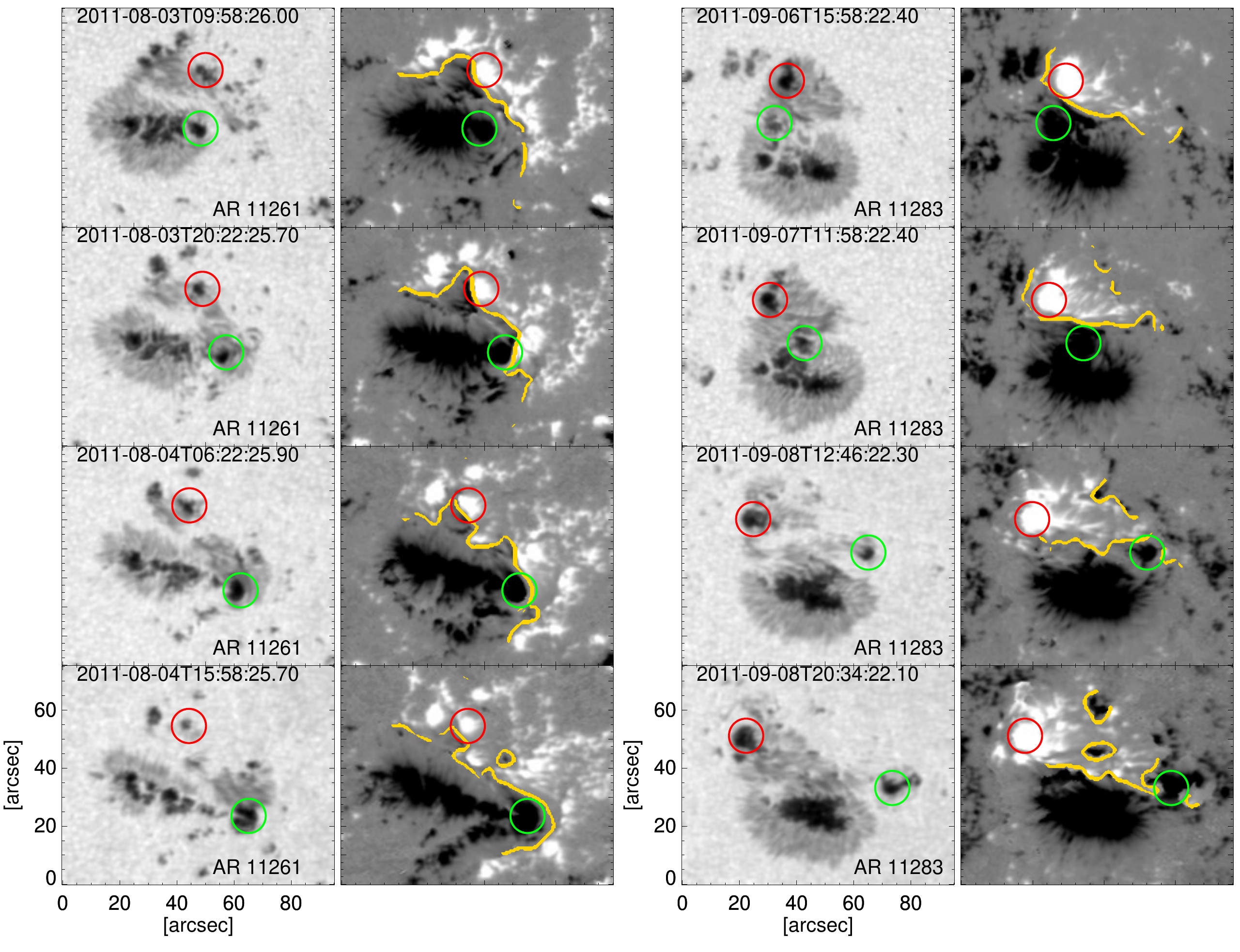}
\caption{HMI continuum images of the flare productive part of AR 11261 (first column) and AR 11283 (third column). Different panels of each column show the temporal evolution of the sunspot group during the recurrent flares. The radial component of the HMI vector magnetic field of AR 11261 (second column) and AR 11283 (fourth column) within the same field of view as shown in the first and third columns respectively. Continuum images in each row of the first/third columns are cotemporal with the magnetic field maps shown in the same row of second/fourth column. The red and green circles depict the two prominent  moving magnetic features of opposite polarities which show continuous antiparallel motion along the polarity inversion line denoted by the yellow solid lines.}
\label{mmf}
\end{figure*}

\begin{figure*}[!t]
\centering
\includegraphics[width=\textwidth]{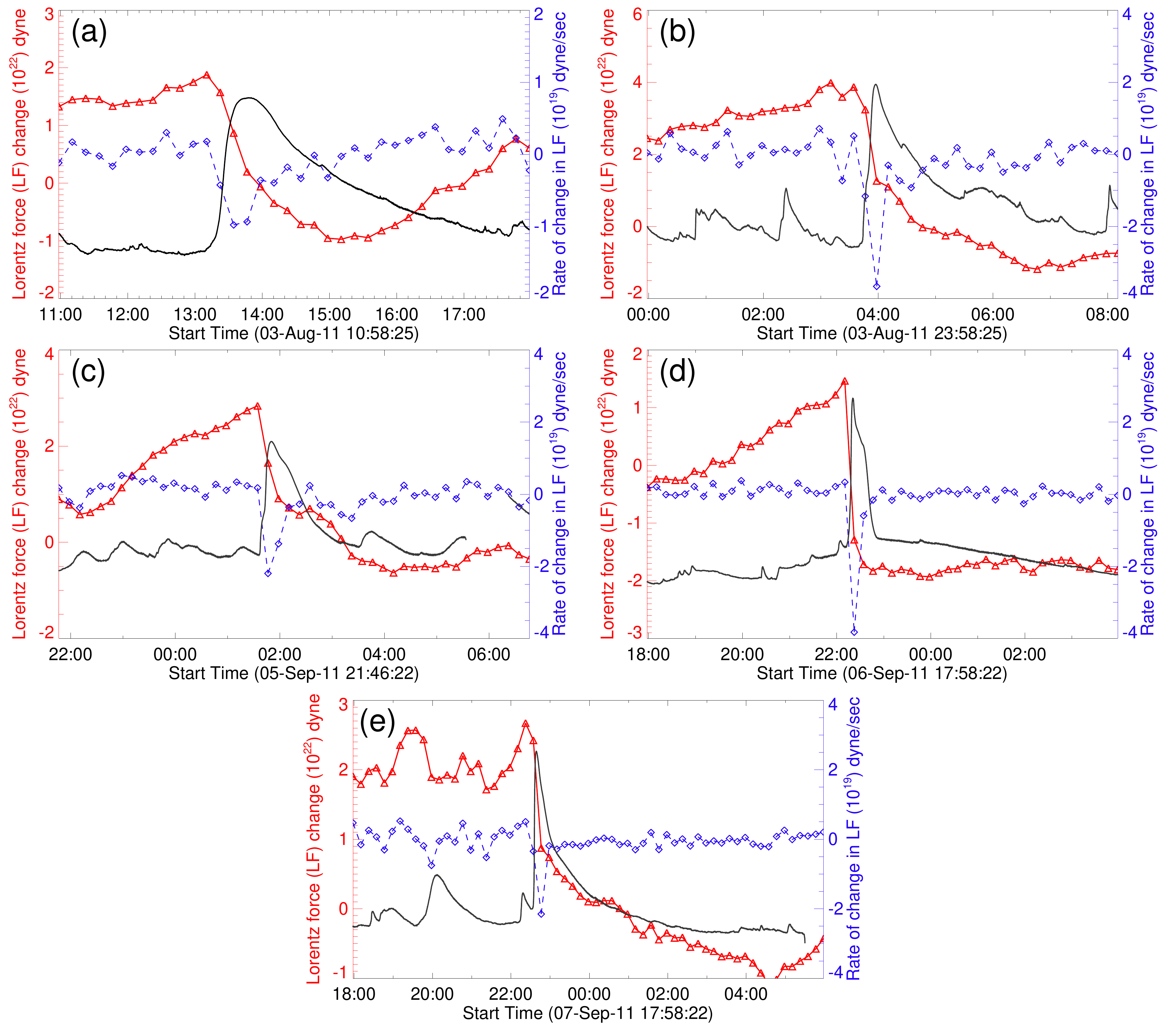}
\caption{Relative evolution of GOES 1-8 \AA\ X-ray flux (black solid lines) with that of the associated Lorentz force (red solid lines) during the recurrent flares under study. The blue dotted line denotes the rate of change in Lorentz force during the flares.}
\label{goes_lf}
\end{figure*}

\subsection{Lorentz Force Rebuild-up in between the Successive Flares}

After the abrupt downward change in net Lorentz force during each large flare that occurred in AR 11261 and AR 11283, the {net} Lorentz force started to rebuild-up in between the successive flares (see Figure \ref{force_plot}). Starting from the magnitude of $-1 \times 10^{22}$ dyne after the M6.0 class, the change in net Lorentz force  reached to a magnitude of $4 \times 10^{22}$ dyne until the next M9.3 class flare occurred in AR 11261. Similarly in AR 11283, the net Lorentz force was rebuilt-up by approximately $2 \times 10^{22}$ dyne in between the M5.3 and X2.1 class flares, and again rebuilt-up by approximately $4 \times 10^{22}$ dyne before the X1.8-class flare. This rebuild-up of the Lorentz force reveals the restructuring of the magnetic field configuration in the vicinity of the PIL in order to increase the non-potentiality of the coronal magnetic field which in turn relaxes by producing the next recurrent flare.

We tested the sensitivity of the obtained results on the size of the bounding boxes selected around the PIL. Increasing the bounding box (see Figure \ref{HMI}) from approx 20 to 40 Mm, the evolutionary pattern of the Lorentz force remains similar. However, integrating the Lorentz force density over the whole AR area dilutes the flare associated changes in the estimated net Lorentz force profile.

The rebuild-up of net Lorentz force in between the recurrent flares could be the consequence of the continuous shearing motion along the PIL. Figure \ref{mmf} shows the continuous shearing motion observed for the two prominent moving magnetic features (MMFs) of opposite magnetic polarities (indicated by the red and green circles). The antiparallel motions of these MMFs along the two sides of the PIL of each AR during the recurrent flares provide evidence for rebuild-up of non-potential energy in between the successive flares. Therefore, the evolution of Lorentz force appears to be a clear indication of energy rebuild-up processes in order to produce successive flares from the same part of any AR. 

Importantly, for the first time we have shown the evolution of a non-potential parameter (net vertical Lorentz force change) that reveals the rebuild-up of non-potentiality of the AR in between the successive large flares. Indeed, this is a significant finding and has important implications. In particular, the evolutionary pattern of the net vertical Lorentz force change can be used for forecasting the recurrent large eruptive flares from the same AR. Furthermore, the associated successive CMEs from the same AR, will in turn enhance their chance of being launched in the same direction. In this scenario, the following faster CME may interact with the preceding slower one in the corona or interplanetary space, which can significantly enhance their geo-effectiveness \citep{Wang2003,Farrugia,Farrugia2,LugazFarrugia}.

Currently available machine-learning algorithms for flare prediction use, among many other parameters, the evolution of Lorentz force integrated over the whole AR, which does not show high skill score in the forecast verification metrics \citep{Bobra_2015}. However, the distinct changes in the vertical component of the Lorentz forces integrated near the PIL demonstrated in our study, could prove to be an important parameter to train and test the machine-learning algorithms in order to improve the current capability of flare-forecasting.

\subsection{Relative Evolution of the GOES X-ray Flux with that of the Associated Lorentz Force During the Flares}
The temporal evolution of the GOES 1-8 \AA\ X-ray flux and the associated change in Lorentz force shows that the Lorentz force starts to decrease at the start of rising phase of the GOES flares (Figure \ref{goes_lf}). Most interestingly, the Lorentz force decreases with a pattern similar to the decay phase of the GOES X-ray flux during all the flares. Among all the five flares (see Table \ref{table1}), the decay phase of the X2.1 class flare (panel (d) of Figure \ref{goes_lf}) was significantly steeper than the other four flares. This reflects in the associated changes in Lorentz force. The Lorentz force also decreases sharply during that X2.1 class flare in comparison to the other flares. The derived rate of change in net Lorentz force associated with the X2.1 class flare is $3 \times 10^{19}$ dyne s$^{-1}$ (Figure \ref{goes_lf}), which is the highest among all the five flares studied in this work. 

These results suggest that the change in Lorentz force is not only related to the phase of impulsive flare energy release, but takes place over a longer interval and follows a similar evolutionary pattern like the decay phase of the GOES soft x-ray flux. This could be associated with a slower structuring of the coronal magnetic field during the decay phase of the flaring events.

\section{\textbf{Conclusion}}\label{third}
Studying the evolution of the photospheric magnetic field and the associated Lorentz force change during the recurrent large flares that occurred in AR 11261 and AR 11283, we find that the vertical component of Lorentz force undergoes abrupt downward changes during all the flares. This result is consistent with earlier studies  \citep{Wang,Petrie2010, Petrie, Sarkar2018}. The observed increase in horizontal magnetic field during each flare is in agreement with the conjecture given by \citet{HFW}, which suggests that the magnetic loops should undergo a sudden shrinkage or implosion due to the energy release processes during flares. This also supports the results obtained by \citet{Romano}, which show a decrease in the dip angle after each large flare that occurred in AR 11283. Interestingly, the decrease in horizontal magnetic field in between the successive flares reported in our study, could be due to the storage of newly supplied energy that increases the coronal magnetic pressure, thereby stretching the magnetic loops upward as  proposed by \citet{Hudson2000}.

Our study also reveals that the decrease in Lorentz force is not only related to the phase of impulsive flare energy release, but takes place over a longer interval that covers also the decay phase of the flaring events. The magnitude of change in net Lorentz forces reported in this work, appears to be correlated with the linear momentum of the associated CME. This scenario is consistent with the flare related momentum balance condition where the Lorentz force impulse is believed to be proportional to the associated CME momentum \citep{Fisher,ShouWang}. 

It is noteworthy that the flare-associated momentum conservation is not only related to the bodily transfer of mass in the form of CMEs, but also includes the effects related to explosive chromospheric evaporation \citep{2012Hudson}. However, quantifying the momentum related to the chromospheric evaporation during the flares under this study is not possible, as this requires spectroscopic observations of both the hot upflowing and cool downflowing plasma. Such measurements are rarely available, due to the localized and dynamic nature of solar flares in contrast to the limited spatio-temporal coverage of spectrometers. However, comparing the values we obtain for the CME momentum, which is of the order of 10$^{15}$ kg km s$^{-1}$, with the momentum related to  chromospheric evaporation flows in large flares as reported in the literature, which is of the order of $10^{13} - 10^{14}$ kg km s$^{-1}$ \citep{1988Zarro,1990Canfield,2012Hudson}, we may conclude that the momentum changes related to the CME are the dominant contribution. Therefore, the correlation between the Lorentz force impulse and the CME momentum  in the large recurrent eruptive flares reported in our study is valid as the  effects of impulsive chromospheric evaporation are at least an order of magnitude smaller.

Most importantly, after the abrupt downward changes during each flare, the net Lorentz force significantly increases to a higher value than that was observed few hours before the flaring event, and only then the subsequent (recurrent) energetic flare occurred. This rebuild-up of net Lorentz force in between the  successive flares suggests that the magnetic field configuration in the vicinity of the PIL is restructured in order to increase the non-potentiality of the coronal magnetic field. Observations of the continuous shearing motions of the MMFs on the two sides of the PIL of each AR provide supporting evidence for rebuild-up of non-potential energy.
 
\citet{Romano} have also reported the shearing motion along the PIL of AR 11283 during the recurrent large M- and X-class flares. They have attributed these photospheric horizontal motions as the possible cause of monotonic injection of magnetic helicity in the corona, which might have resulted in the episodic energy release processes, leading to the recurrent flares. However, the evolution of the horizontal magnetic field and the  associated Lorentz force reported in our study, clearly indicates the energy rebuild-up processes in order to produce successive flares from the same part of the AR. Therefore, we conclude that the recurrent flares studied in this work occurred due to the newly supplied energy to the AR through the continuous shearing motions of photospheric magnetic field in between the successive flares. 

We thank the referee for helpful comments that improved the quality of this manuscript. We acknowledge NASA/SDO and the AIA and HMI science teams for data support. A.M.V. acknowledges the Austrian Science Fund (FWF): P27292-N20. This work was supported by the Indo-Austrian joint research project no. INT/AUSTRIA/BMWF/P-05/2017 and OeAD project no. IN 03/2017.

\bibliographystyle{yahapj}

\end{document}